# Nodal superconducting exchange coupling

A. Di Bernardo[1†*], S. Komori[1], G. Livanas[2], G. Divitini[1],
P. Gentile[2], M. Cuoco[2], J.W.A. Robinson[1*]

*Corresponding author. Email: angelo.dibernardo@uni-konstanz.de (A.D.B.)
and jjr33@cam.ac.uk (J.W.A.R.)

1. Department of Materials Science & Metallurgy, University of Cambridge, 27 Charles Babbage Road, Cambridge CB3 0FS, United Kingdom.

2. CNR-SPIN, c/o University of Salerno, I-84084 Fisciano, Salerno, Italy.

†Current address: Department of Physics, University of Konstanz, Universitätstraße 10, 78457 Konstanz, Germany.

The superconducting equivalent of giant magnetoresistance[1,2] involves placing a thin-film superconductor between two ferromagnetic layers. A change of magnetization-alignment in such a superconducting spin-valve from parallel (P) to antiparallel (AP) creates a positive shift in the superconducting transition temperature ($\Delta T_c$) due to an interplay of the magnetic exchange energy and the superconducting condensate. The magnitude of $\Delta T_c$ scales inversely with the superconductor thickness ($d_S$) and is zero when $d_S$ exceeds the superconducting coherence length ($\xi$) as predicted by de Gennes[3]. Here we report a superconducting spin-valve effect involving a different underlying mechanism that goes beyond de Gennes in which magnetization-alignment and $\Delta T_c$ are determined by the nodal quasiparticle-excitation states on the Fermi surface of the *d*-wave superconductor $YBa_2Cu_3O_{7-\delta}$ (YBCO) grown between insulating layers of ferromagnetic $Pr_{0.8}Ca_{0.2}MnO_3$. We observe $\Delta T_c$ values that approach 2 K with $\Delta T_c$ oscillating with $d_S$ over a length scale exceeding 100 $\xi$ and, for particular values of $d_S$, we find that the superconducting state reinforces an antiparallel magnetization-alignment. These results pave the way for all-oxide superconducting memory in which superconductivity modulates the magnetic state.

Predating the discovery of giant magnetoresistance[1,2], in 1966 de Gennes[3] proposed a magnetic memory concept in which the superconducting transition temperature ($T_c$) of a thin-film *s*-wave superconductor (S) between ferromagnetic insulators (FIs) is dependent on magnetization alignment. The underlying mechanism controlling $\Delta T_c$ relates to the magnetic exchange field at the S/FI interfaces inducing a spin-splitting in the (gapped) density of states (DoS) in S[4,5], which decays over a superconducting coherence length ($\xi_S$). Exchange fields at S/FI interfaces are additive meaning that the largest (smallest) net exchange field acting on S corresponds to a P (AP) magnetization alignment of the FIs[6,7] - i.e. $\Delta T_c = T_c$ (P) - $T_c$ (AP) is negative[6,7,8,9,10,11,12,13,14], although positive values have been reported due to spin-accumulation[15,16,17] or magnetic stray fields suppressing $T_c$ in the AP-state coupling through S[18,19,20,21,22].

Over the past several decades, different types of superconducting memory devices have been investigated involving *s*-wave S in combination with FIs (e.g. EuS[6] and GdN[7]), transition metal ferromagnets[8,13,15,16,17,18] and rare earths ferromagnets[14]. In all these experiments, except those dominated by spin-accumulation[15] or magnetic stray fields[18], $\Delta T_c$ falls to zero as the thickness of the S layer exceeds $\xi_S$ in agreement with de Gennes theory[3].

Equivalent superconducting spin-valves involving half-metallic ferromagnets[16,17,19,20,21,22] have also been investigated in combination with the nodal (*d*-wave) superconductor YBa$_2$Cu$_3$O$_{7-\delta}$ (YBCO). In YBCO, the coherence length is anisotropic with $\xi_{ab} \approx 1.6$ nm and $\xi_c \approx 0.2 - 0.3$ nm[23] in the *ab*-plane and *c*-axis directions, respectively. In these experiments, $\Delta T_c$ tends to be dominated by magnetic stray fields[16,17] or spin-accumulation in YBCO[19,20,21,22] rather than by the proximity effect. Recently, $\Delta T_c$ has been investigated[24] in superconducting spin-valves involving the electron-doped superconductor Pr$_{1.85}$Ce$_{0.15}$CuO$_4$ (PCCO; $\xi_{ab} \approx 6$-$8$ nm and $\xi_c \approx 1$ nm[25]) sandwiched between the FI Pr$_{0.8}$Ca$_{0.2}$MnO$_3$ (PCMO). Here a genuine proximity effect was observed, although $\Delta T_c$ was less than 50 mK and was localized to the PCCO/PCMO interface[24]. Unlike YBCO, however, electron-doped PCCO does not show evidence for nodal quasiparticle states meaning that the superconducting gap is isotropic[26,27].

In this paper, we report PCMO/YBCO/PCMO devices and demonstrate three key results: firstly, $\Delta T_c$ approaches 2 K; secondly, the sign and magnitude of $\Delta T_c$ oscillate with YBCO thickness ($d_S$) over a length scale that is two orders of magnitude larger than $\xi_c$; and thirdly, for particular values of $d_S$ the superconducting state reinforces AP coupling. The latter two cannot be explained in the context of de Gennes theory[3]. In conjunction with a theoretical model, we demonstrate that our results can be explained in terms of a new proximity effect in which the PCMO layers couple through YBCO via the nodal quasiparticle states in the *c*-axis direction. Control experiments rule out any influence due to magnetic stray fields from the PCMO layers affecting $\Delta T_c$ and spin-accumulation is not relevant since PCMO is an insulator.

A series of epitaxial PCMO(50 nm)/YBCO($d_S$)/PCMO(100 nm) trilayers are grown by pulsed laser deposition onto single-crystal (001) SrTiO$_3$ (STO), each with a $d_S$ value of 4, 5, 7, 9, 11, 12 and 17 unit cells (u.c.) (see **Methods**). High-resolution transmission electron microscopy (TEM) in Fig. **1a**, **b** and high-angle X-ray diffraction (Fig. **1c** and Extended Data Fig. **1**) confirm epitaxial growth of the trilayers with atomically-sharp interfaces (see insets to Fig. **1b**) and the *c*-axis oriented out-of-plane. This is further confirmed by additional TEM and high-resolution Scanning TEM (HR-STEM) maps acquired on our trilayers and reported in Extended Data Fig. **2a**-**c**. STEM Electron Dispersive X-ray (STEM-EDX) maps demonstrate the absence of intermixing between layers (Extended Data Fig. **2d**-**i**). We use Principal Component Analysis and Non-negative Matrix Factorization on the STEM-EDX spectra with a pixel size < 0.5 nm[28], which rules out the presence of additional phases other than those expected or any changes in local stoichiometry at the YBCO/PCMO interfaces (Extended Data Fig. **3** and **4**). Electron-Energy Loss Spectroscopy (EELS) maps acquired on the samples show that the Mn oxidation state is constant at the YBCO/PCMO interfaces as well as across the PCMO layers thus ruling out charge transfer effects between PCMO and YBCO (Extended Data Fig. **5**).

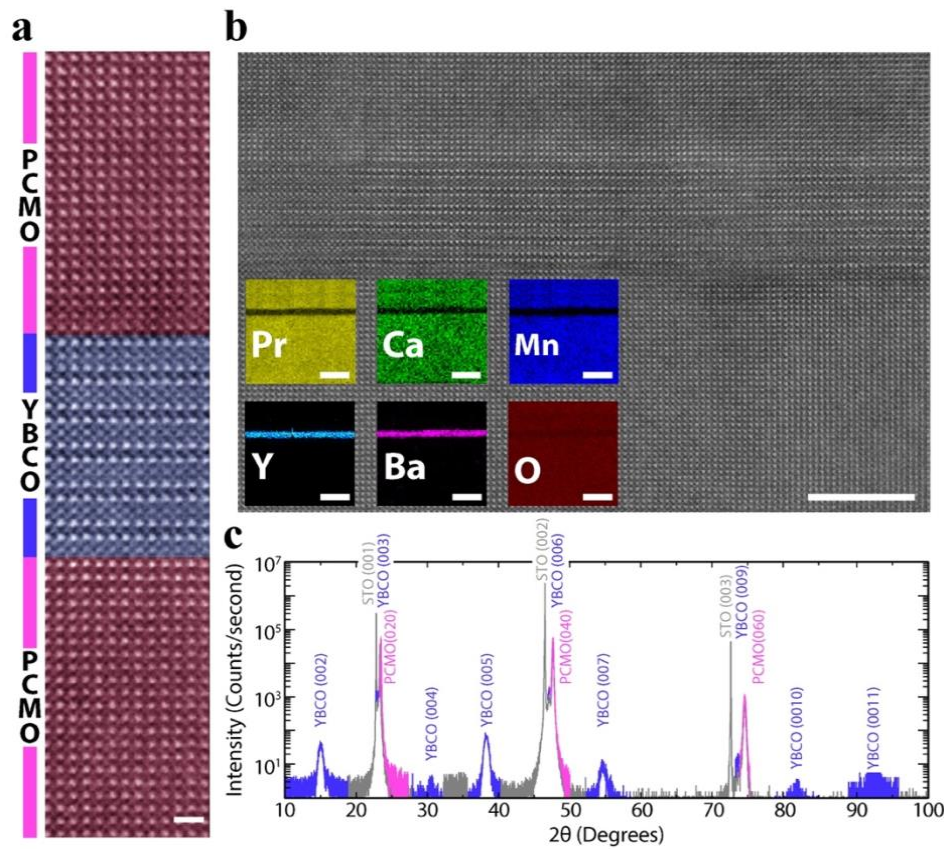

**Fig. 1: Structure of PCMO(50 nm)/ YBCO(5 u.c.)/ PCMO(100 nm).** **a**, **b**, High-resolution scanning transmission electron micrographs with insets in **b** showing electron energy-loss compositional maps for Pr (yellow), Ca (green), Mn (blue), Y (light blue), Ba (magenta), and O (burgundy). The scale bar in **a** corresponds to 1 nm whilst in **b**, including insets, to 10 nm. **c**, High-angle X-ray diffraction data from the same sample confirming *c*-axis growth out-of-plane.

The electronic resistance ($R$) of YBCO between PCMO layers is investigated as a function of $d_S$, temperature ($T$), and external magnetic field ($H$) directed parallel to the YBCO/PCMO interfaces. A four-point electrical setup is used with electrical contact directly to YCBO (Fig. **2a**) otherwise electrical currents cannot pass through the insulating (top) layer of PCMO (see refs 29, 30 and Extended Data Fig. **6**). A typical set of $R(T)$ plots around the superconducting transition are shown in Fig. **2a, b** for trilayers with $d_S = 17$ u.c. and $d_S = 9$ u.c. In Extended Data Fig. **7a, b** we have plotted example $R(T)$ curves along with the temperature width of the superconducting transition and $T_c$ at zero resistance ($T_{c,0}$) versus $d_S$ for a series of PCMO(50 nm)/YBCO($d_S$)/PCMO(100 nm) trilayers. These results show that $T_c$ decreases monotonically with $d_S$ with a critical YBCO thickness of 4 u.c. needed to achieve a full superconducting transition, consistent with related studies[31,32].

We first discuss $R(H)$ curves through the superconducting transition. In Figure **2** we have plotted a set of $R(H)$ curves for two trilayers with $d_S = 17$ u.c. (Fig. **2c-e**) and $d_S = 9$ u.c. (Fig. **2i-k**). Corresponding magnetization versus in-plane magnetic field ($M(H)$) loops for the same trilayers at matching temperatures are also plotted in Figure **2**: $d_S = 17$ u.c. in Fig. **2f-h**, and $d_S = 9$ u.c. in Fig. **2l-n**. Equivalent $R(H)$ curves are plotted in Extended Data Fig. **8** for $d_S$ values of 5 u.c., 7 u.c. and 12 u.c.

For all values of $d_S$ investigated, $R(H)$ shows magnetoresistance (MR) through the superconducting transition with well-defined changes in $R$ corresponding to the positive and negative coercive field ($H_c$) values of the PCMO layers. The MR originates from changes in $T_c$ between P and AP magnetization-alignments of the PCMO layers with the sign and magnitude of $\Delta T_c$ oscillating with $d_S$. We note that, for all trilayers investigated, $R(H)$ shows no measurable dependence on magnetization-alignment above $T_c$ meaning that changes in $R(H)$ are related to superconductivity and not conventional (non-superconducting) magnetoresistance.

Figure **2** highlights two fundamentally different $R(H)$ behaviours. For both trilayers in Fig. **2**, $R(H)$ shows a local minimum in $R$ in the AP-state ($R_{AP}$) for all values of $T$ through the superconducting transition; however, the $H$-position of the absolute minimum in $R$ ($R_{min}$) does not necessarily correspond to the AP-state: for $d_S = 17$ u.c., $R_{min}$ occurs in the AP-state (i.e. $R_{min} = R_{AP}$), but for $d_S = 9$ u.c., $R_{min}$ occurs in the P-state (i.e. $R_{min} = R_P$). Hence, for $d_S = 17$ u.c., $T_c$ is higher in the AP-state (i.e. $\Delta T_c < 0$) while for $d_S = 9$ u.c., $T_c$ is higher in the P-state (i.e. $\Delta T_c > 0$). These opposite behaviours are systematically observed in all 7 sets of trilayers investigated (see also Extended Data Fig. **8**).

Although $R_{min}$ always coincides with either $R_{AP}$ or with $R_P$, we observe that the maximum resistance value ($R_{max}$) is mostly measured at $H \sim 0$ (if the $R$ suppression due to $H$ is taken into account). The $M(H)$ loops of the PCMO/YBCO/PCMO trilayers (Fig. **2** and Extended Data Figs. **9** and **10**) show a reduced magnetization in the remanent state ($H = 0$), which indicates an inhomogeneous micromagnetic state within and between the PCMO layers. The correspondence between $R_{max}$ with this inhomogeneous magnetic state of the FIs at $H \sim 0$ is further discussed in our theoretical analysis below.

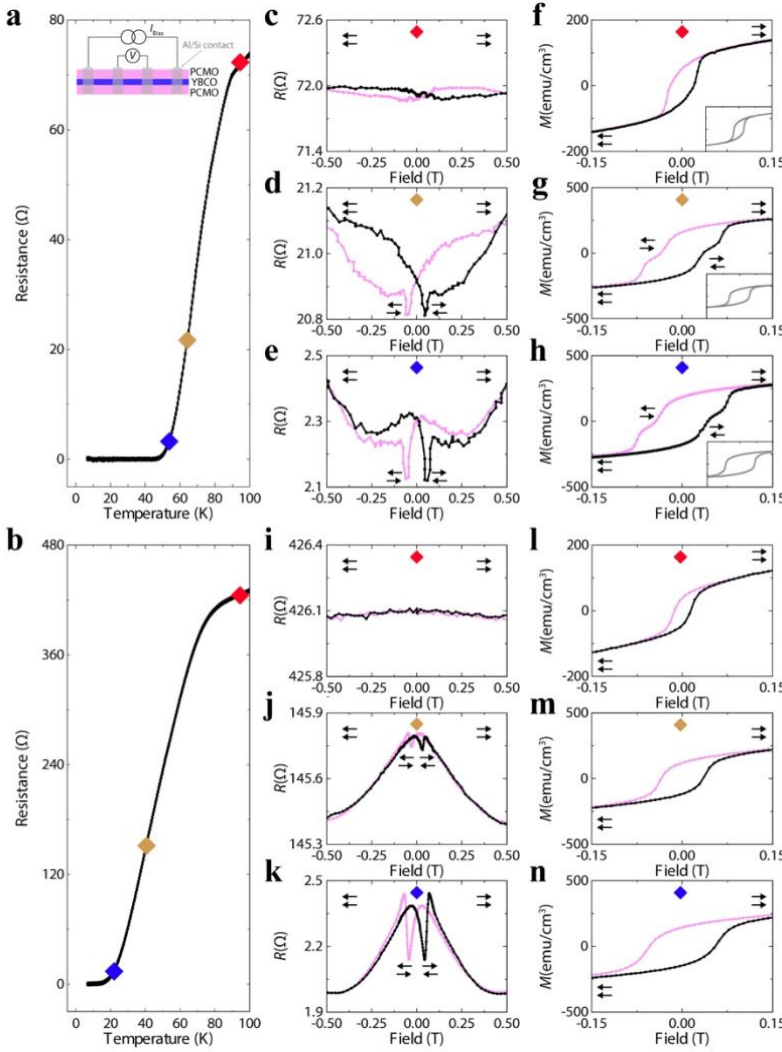

**Fig. 2: Electronic and magnetic properties of PCMO/YBCO/PCMO trilayers. a**, **b**, $R(T)$ across the superconducting transition for PCMO(50 nm)/YBCO($d_S$)/PCMO(100 nm) trilayers with $d_S = 17$ u.c. in **a** and 9 u.c. in **b**; **c-h**, $R(H)$ curves (**c-e**) and $M(H)$ loops (**f-h**) measured at different temperatures across the superconducting transition for the $d_S = 17$ u.c. trilayer with the diamond symbol (♦) in **c-h** specifying the temperatures marked in **a** using matching symbols; **i-n**, $R(H)$ curves (**i-k**) and $M(H)$ loops (**l-n**) for the $d_S = 9$ u.c. trilayer with the diamond symbols (♦) in **i-n** specifying the temperatures marked in **b** using matching symbols. Insets in **f-h** show $M(H)$ loops for PCMO(50 nm)/PCO(5 nm)/YBCO(17 u.c.)/PCO(5 nm)/PCMO(100 nm) control samples in which the proximity effect between PCMO and YBCO is blocked via the non-magnetic insulator PCO. The arrows indicate the magnetization orientation of the FI layers.

An additional important observation is that in all trilayers ($d_S = 7, 11, 12,$ and $17$ u.c.) in which the superconductivity favours the AP-state (i.e. $R_{min}$ occurs in the AP-state), $M(H)$ shows an increasing mismatch in $H_c$ ($\Delta H_c$) between the top and bottom PCMO layers as $T$ is lowered through the superconducting transition. This is demonstrated in Fig. **2f-h** and in Extended Data Fig. **9a** for the $d_S = 17$ u.c. trilayer which shows an increasingly-wide field range for the AP-state below $T_c$. Equivalent control samples in which PCMO and YBCO are decoupled with a 5-nm-thick film non-magnetic insulator $Pr_2CuO_4$ (PCO) exhibit no detectable $\Delta H_c$ for all temperatures above and below $T_c$ as shown in the insets to Fig. **2f-h** and in Extended Data Fig. **9c**. A similar increase in $\Delta H_c$ to the trilayer with $d_S = 17$ u.c. is also observed in the $M(H)$ loops measured for the trilayers with $d_S = 7$ u.c. (Extended Data Fig. **10a**) and $d_S = 12$ u.c. (Extended Data Fig. **10b**).

For trilayers ($d_S$ = 4, 5 and 9 u.c.) in which the P-state is favoured below $T_c$ (i.e. $R_{min}$ occurs in the P-state), $\Delta H_c$ remains negligible for all $T$ as shown in Fig. **2l-n** and Extended Data Fig. **9b** for the case of $d_S$ = 9 u.c.

These observations demonstrate that the magnetic configuration between the PCMO layers is dependent on $d_S$ and the superconducting state in YBCO, since $\Delta H_c$ is correlated with the sign of $\Delta T_c$ - i.e. for devices with $\Delta T_c < 0$ ($T_c$(AP) > $T_c$(P)), $\Delta H_c$ increases below $T_c$, whilst for devices that show $\Delta T_c > 0$ (i.e. $T_c$(P) > $T_c$(AP)), $\Delta H_c$ is zero for all $T$. This is summarised in Fig. **3a-f**, where the $T$-evolution of $\Delta H_c$ with $R(T)$ is shown through the superconducting transition for different $d_S$.

Finally, for all values of $d_S$ studied, we investigate $R(T)$ in the P and AP states. In Fig. **3g** we have plotted the maximum values of $\Delta T_c$ as a function of $d_S$ which show an oscillatory behaviour with $\Delta T_c$ switching between positive and negative values, reaching a maximum of 1.7 K.

The long-ranged dependence of $\Delta T_c$ on $d_S$ indicates physics beyond de Gennes in that significant values of $\Delta T_c$ are measurable with $d_S$ up to 17 u.c., which corresponds to approximately 100 $\xi_c$ for YBCO. This explanation necessarily implies that the magnetic exchange field at the PCMO/YBCO interfaces is able to penetrate YBCO through the states close to the nodal points on the Fermi surface for which the effective superconducting coherence length can be large due to the vanishing pairing gap[33].

Alternative explanations could involve out-of-plane magnetic stray fields from PCMO suppressing $T_c$ at $H_c$ or spin-accumulation in YBCO creating a spin-imbalance and suppression of $T_c$ in the AP state as observed in YBCO/La$_{0.66}$Ca$_{0.33}$MnO$_3$ in refs [19,20,21,22]. In Extended Data Fig. **11** we have plotted $R(H)$ curves and $M(H)$ loops measured at matching temperatures across the superconducting transition for control samples consisting of YBCO($d_S$)/PCMO(100 nm) bilayers with $d_S$ = 6 u.c. and 17 u.c. as well as of PCMO(50 nm)/YBCO($d_S$) bilayers with $d_S$ = 12 u.c. and 18 u.c, which show no fluctuations in $R(H)$ around $H_c$ of PCMO thus ruling out a stray field suppression of $T_c$ due to potential out-of-plane Bloch domain walls. Furthermore, spin-accumulation in YBCO is not possible due to the fact that PCMO is an insulator meaning that charge cannot flow from PCMO into YBCO. The matching in $H_c$ and in magnetotransport behaviour as a function of $T$ between PCMO(50 nm)/YBCO bilayers and YBCO/PCMO(100 nm) bilayers shown in Extended Data Fig. **11** also further demonstrates evidence for the absence of antiferromagnetic-like coupling between PCMO and YBCO at PCMO/YBCO or PCMO/YBCO interfaces. Antiferromagnetic coupling between PCMO layers in PCMO/YBCO/PCMO trilayers is therefore indeed mediated via YBCO in its superconducting state. We also note that the shape and size of the dips and peaks in $R(H)$ do not depend on the in-plane current direction relative to $H$ (Extended Data Fig. **12**).

A long-ranged effect of the magnetic exchange energy ($E_{ex}$) in YBCO in combination with the fact that $\Delta H_c$ in the superconducting state is $d_S$-dependent suggests that the PCMO layers are magnetically coupled through YBCO via its states close to the nodal points. The presence of an $E_{ex}$ in YBCO will energy-split its Dirac-like quasiparticle excitation spectrum and lead to the formation of Fermi pockets as sketched in Figure **3h**. As a result, YBCO is no longer a nodal ($d$-wave) S with $N$ nodal points at the Fermi surface ($N$ being the number of u.c.), but one which contains $N$ Fermi pockets at the Fermi surface. This is explained in detail in the **Supplementary Information**.

In Fig. **3i** we show the Fermi pockets in YBCO in the two-dimensional Brillouin zone (BZ) for $d_S = 6$ u.c. In the **Supplementary Information** we have calculated the Gibbs free energy ($\mathcal{F}$) of the entire system at different $k$-channels and integrated all the $k$-channel contributions over the entire BZ taking into account the pairing interaction energy, temperature, $E_{ex}$, and the intra- and inter-u.c. hopping of charge. It is found that $\mathcal{F}$ for a given $k$-channel, $\tilde{\mathcal{F}}(\theta)$, is dependent on the magnetization angle ($\theta$) between the PCMO layers (see insets to Fig. **3i**) showing a minimum at $\theta = 0$ (P state) or $\pi$ (AP state) depending on the $k$-channel position with respect to the Fermi pockets. This has the consequence that global minimum in $\mathcal{F}(\theta)$ of PCMO/YBCO/PCMO at a given $d_S$, which is obtained by summing up the $\tilde{\mathcal{F}}(\theta)$ contributions of all $k$-channels, can manifest a competition and prefer either a P or an AP state.

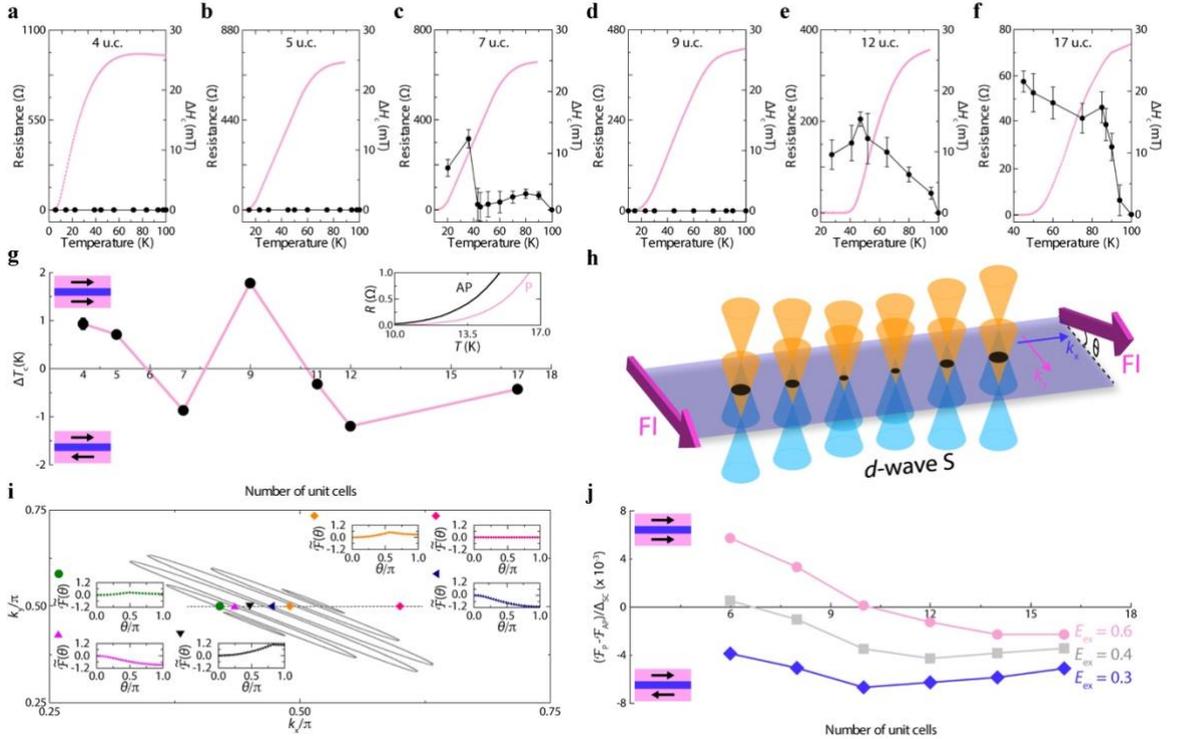

**Fig. 3: Exchange coupling in PCMO/YBCO/PCMO. a-f**, $R(T)$ curves showing $\Delta H_c$ through the superconducting transition for trilayers with different values of $d_S$ (labelled); **g**, maximum values of $\Delta T_c = T_c$ (P) - $T_c$ (AP) versus $d_S$ with the inset showing $R(T)$ curves in the P and AP states for $d_S = 9$ u.c; **h**, a sketch illustrating the energy splitting of low-energy quasiparticle excitations in YBCO due to $E_{ex}$ at the YBCO/FI interfaces; **i**, calculated Fermi surface of YBCO between two FIs with magnetization angle $\theta$. Insets show free energy curves at points in $k$-space versus $\theta$. **j**, Minima in free energy for P

(top) and AP (bottom) states versus $d_S$ for different values of $E_{ex}$ (given in units of the intra-u.c. charge hopping parameter). The sketches in **g** and **j** show the ground state (P or AP) of the trilayer.

In addition, by varying $d_S$ the contribution of each $k$-channel to $\mathcal{F}(\theta)$ changes since the number and size of the Fermi pockets is $d_S$-dependent. Consequently, YBCO will set a preferred magnetization alignment (P or AP) which is also $d_S$-dependent, consistently with our experimental results in Fig. **3g**. We note that although the coupling between the PCMO layers mediated by YBCO has an oscillatory dependence on $d_S$, this exchange coupling is fundamentally different from a Ruderman-Kittel-Kasuya-Yosida (RKKY) coupling[34], as explained in the **Supplementary Information**.

The insets in Fig. **3i** also show that $\tilde{\mathcal{F}}(\theta)$ can have either a single absolute minimum at $\theta = 0$ (black curve) or at $\theta = \pi$ (blue and magenta curves) or two minima (one absolute and one relative) at $\theta = 0$ and $\theta = \pi$ (orange and green curves). In the latter case, $\tilde{\mathcal{F}}(\theta)$ reaches a maximum value for $0 < \theta < \pi$, corresponding to a magnetically-inhomogeneous state for the system (Extended Data Fig. **9** and **10**). Such trends of $\tilde{\mathcal{F}}(\theta)$ are thus consistent with our experimental $R(H)$ data showing that $R(H)$ has either a single absolute minimum (e.g. $d_S = 17$ u.c. in Fig. **2d, e** and $d_S = 7$ u.c. in Extended Data Fig. **8d**) or two minima at $\theta = 0$ and $\theta = \pi$ (e.g. $d_S = 9$ u.c. in Fig. **2j, k** and $d_S = 5$ u.c. in Extended Data Fig. **8b**) with $R_{max}$ measured at $H \sim 0$, where the FIs are in a non-collinear state. Our theoretical analysis demonstrates that the global $\mathcal{F}(\theta)$ of the FI/S/FI system can also follow a similar trend to $\tilde{\mathcal{F}}(\theta)$, although depends in a complex way on $E_{ex}$ and $H$. When $\mathcal{F}(\theta)$ has two minima at $\theta = 0$ and $\theta = \pi$, a maximum is obtained for $0 < \theta < \pi$ which shifts towards $\theta = 0$ or $\theta = \pi$ at $H = 0$ depending on $E_{ex}$. An applied $H$ instead, for a fixed $E_{ex}$ value, changes the absolute minimum in $\mathcal{F}(\theta)$ from $\theta = \pi$ to $\theta = 0$, while the maximum in $\mathcal{F}(\theta)$ progressively shifts towards $\theta = \pi$ (see **Supplementary Information**).

Our Gibbs free energy analysis also predicts that the preferred magnetization alignment of PCMO/YBCO/PCMO should depend on the magnitude of $E_{ex}$ and indirectly therefore on the Curie temperature ($T_{Curie}$) of PCMO (see **Supplementary Information**). Furthermore, below a critical value of $E_{ex}$ the AP state is theoretically preferred for all values of $d_S$ (Fig. **3j**). To test this experimentally, we fabricate two series of trilayers with $d_S = 12$ u.c. and 5 u.c. in which the Ca doping ($x$) of the Pr$_{1-x}$Ca$_x$MnO$_3$ is symmetrically varied with $x = 0.2$ ($T_{Curie} \approx 130$ K), 0.1 ($T_{Curie} \approx 110$ K) and 0.05 ($T_{Curie} \approx 90$ K). The magnetic properties of these Pr$_{1-x}$Ca$_x$MnO$_3$ films with different $x$ are reported in Extended Data Fig. **13**. In Extended Data Fig. **14** ($d_S = 12$ u.c.) and Extended Data Fig. **15** ($d_S = 5$ u.c.) we have plotted a series of $R(H)$ curves which confirm that the AP-state is preferred for $x = 0.05$ independently on $d_S$ and that, for $x = 0.2$ and 0.1 instead, the preferred magnetization state remains $d_S$-dependent.

The experimental results in conjunction with our theoretical model demonstrate that $\Delta T_c$ and magnetization-alignment depend on an interaction between the magnetic exchange energy and the nodal states in YBCO. In a related experiment[24] we substituted YBCO for the electron-doped superconductor $Pr_{1.85}Ce_{0.15}CuO_4$ (PCCO) in which nodal states do not exist due to an averaging out of the DoS meaning that PCCO is fully gapped in all directions in $k$-space[26,27]. For PCMO/PCCO/PCMO trilayers, changes in $T_c$ due to the underlying magnetization direction of the PCMO layers were due to a proximity effect in PCCO that occurred over a distance of the order of $\xi_c \approx 1$ nm[25] from the PCCO/PCMO interface thus following the behaviour predicted by de Gennes.

Finally, we note that in all previous F/S/F (where F is a ferromagnet) and FI/S/FI experiments ("Superconducting spin valves") with metallic- or oxide materials, the sign of $\Delta T_c$ is constant with $d_S$ (Fig. **4**). For PCMO/YBCO/PCMO trilayers, however, we observe a sign change in $\Delta T_c$ with $d_S$, with $\Delta T_c$ approaching 2 K. These values are consistent also with those we expect from pure theoretical considerations (see **Supplementary Information**). Furthermore, for certain values of $d_S$, the presence of superconductivity in YBCO reinforces an AP-state depending on the number of available Fermi pockets through which the exchange fields from PCMO can couple.

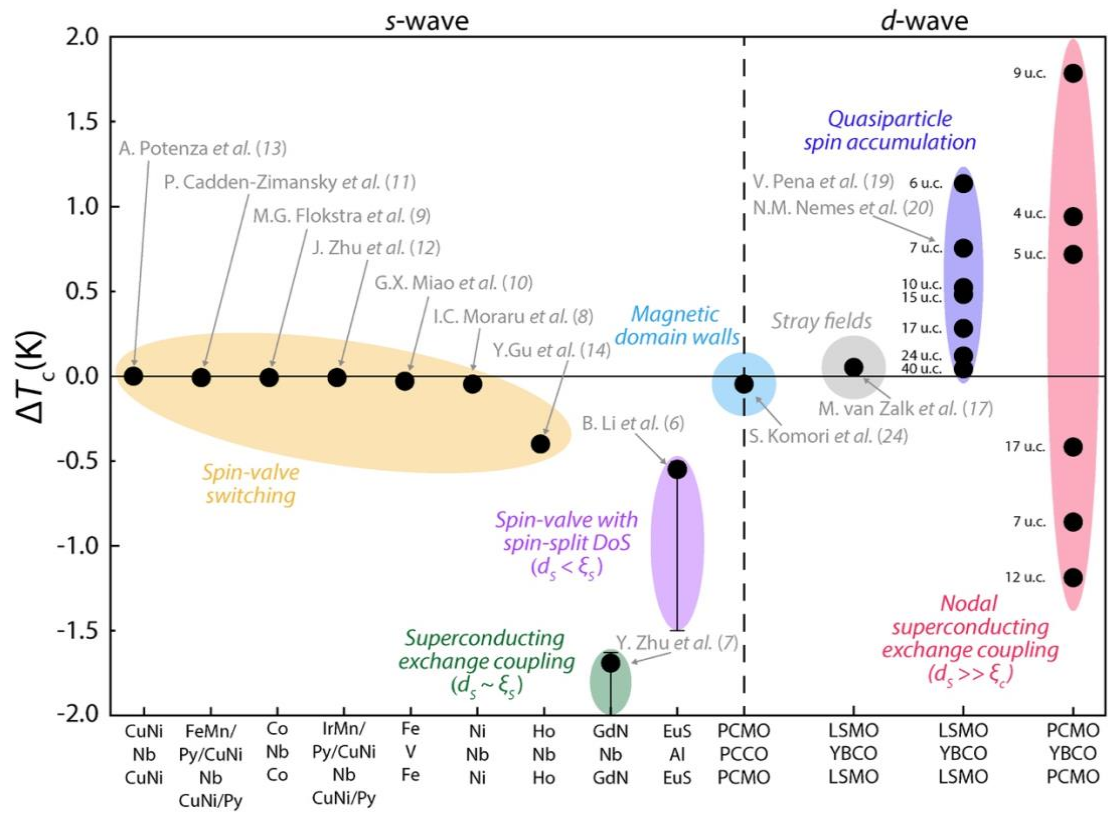

**Fig. 4: Different types of superconducting spin-valves.** $\Delta T_c$ values for F/S/F and FI/S/FI trilayers using different materials combinations involving metallic ($s$-wave) and oxide ($d$-wave) superconductors with ferromagnets (Fs) and ferromagnetic insulators (FIs). $\Delta T_c$ values are clustered according to the underlying mechanism (labelled) which controls the magnitude and sign of $\Delta T_c$.

## Methods

**Sample growth**.

$Pr_{1-x}Ca_xMnO_3$(50 nm)/YBCO($d_S$)/$Pr_{1-x}Ca_xMnO_3$(100 nm) trilayers with $x = 0.05, 0.1$ and $0.2$ and variable $d_S$ are grown in an ultra-high vacuum pulsed laser deposition (PLD) system onto (001) $SrTiO_3$ substrates (CrysTec GmbH) from stoichiometric targets of $Pr_{1-x}Ca_xMnO_3$ fabricated by solid-state reaction from $Pr_6O_{11}$ (99.9% purity), $CaCO_3$ (99.95% purity) and $MnO_2$ (99.9% purity) powders and a commercial YBCO target (99.99% purity; MTI corporation). All layers are grown at 780 ºC in 200 mTorr of $O_2$ using a Lambda Physik KrF excimer laser (wavelength of 248 nm) with an energy density of 1.5 J/cm$^2$. The targets are rotated during growth and the $Pr_{1-x}Ca_xMnO_3$ target is ablated at a pulse rate of 2 Hz and YBCO at 1 Hz. Following growth, trilayers are rapidly-cooled in $O_2$ to 500 ºC and annealed at this temperature in 500 Torr of $O_2$ for 1 hour before being cooled down to room temperature at the same $O_2$ pressure.

**Transport and magnetic measurements.**

Electrical resistance is measured in a pulse tube cryogen-free system (Cryogenic Ltd) with an in-plane magnetic field and temperature stability of less than 10 mK. The current-voltage characteristics are measured using a current-bias of equal to or less than 0.1 mA using a four-probe configuration with electrical contacts directly to YBCO. The contacts are fabricated through an optimised ultrasonicating bonding protocol with AlSi wire. We find that if the right parameters are not used during bonding, connections are made to the top PCMO layer resulting a very large $R$ with an insulating trend as a function of $T$. Magnetization hysteresis loops are measured using a magnetic property measurement system (Quantum Design Ltd) with a moment sensitivity of better than $10^{-8}$ emu.

**Transmission electron microscopy (TEM).**

A lamella for TEM analysis is prepared using a FEI Helios Nanolab focused ion beam/scanning electron microscope (FIB/SEM). The process involves depositing a 2-µm-thick capping layer of Pt, liftout of the region of interest and thinning down to electron transparency (< 100 nm). Scanning TEM STEM-HAADF (High-Angle Annular Dark Field) images are acquired on a probe-corrected FEI Titan operated at 300 kV. Electron Energy-Loss Spectroscopy (EELS) maps are acquired on a FEI Osiris (200 kV acceleration voltage) equipped with a high-brightness field emission gun (X-FEG) and a Gatan Enfinium EELS spectrometer. EELS data are analysed using Gatan Digital Micrograph 2.33. STEM-EDX data are acquired on the same instrument with a Bruker Super-X detector, comprising 4 silicon drift detector quadrants with a total acquisition angle of 0.9 sr. The EDX maps, with a spatial sampling of 0.5 nm, are analysed using Principal Component Analysis (PCA) and Non-negative Matrix Factorization (NMF) through an open source python-based toolkit for electron microscopy data analysis (Hyperspy). High-resolution STEM Energy Dispersive X-ray (HRSTEM-EDX) data are acquired on a JEOL GrandARM operated at 300 kV, equipped with a cold field emission gun and two silicon-drift EDX detectors, courtesy of JEOL's factory in Akishima, Japan, and processed using JEOL's proprietary software into elemental maps.

## Acknowledgments

A. D. B., S. K. J. W. A. R. acknowledge funding from the EPSRC through the International Network Grant No. EP/N017242/1. A. D. B. acknowledges funding from St John's College, Cambridge. J. W. A. R. acknowledges funding from the Royal Society.

## Author information

### Affiliations

*Department of Materials Science & Metallurgy, University of Cambridge, Cambridge, UK*
Angelo Di Bernardo, Sachio Komori, Giorgio Divitini & Jason W. A. Robinson

*CNR-SPIN, c/o University of Salerno, Salerno, Italy*
Georgios Livanas, Paola Gentile & Mario Cuoco.

### Contributions
J.W.A.R. supervised the project and with A.D.B. conceived and designed it. A.D.B. prepared the samples and performed electrical and magnetic measurements with help from S. K. TEM imaging and compositional analysis were performed by G. D. The theoretical model was developed by M.C and P. G. with support from G. L, and input from A.D.B. and J.W.A.R. The paper was written by A.D.B., J.W.A.R. and M.C. with input from all authors.

### Corresponding authors
Correspondence to Angelo Di Bernardo (angelo.dibernardo@uni-konstanz.de) or Jason W. A. Robinson (jjr33@cam.ac.uk)

# Supplementary Information for

## Nodal superconducting exchange coupling

A. Di Bernardo[1†*], S. Komori[1], G. Livanas[2], G. Divitini[1], P. Gentile[2], M. Cuoco[2],

J. W.A. Robinson[1*]

1. Department of Materials Science & Metallurgy, University of Cambridge, 27 Charles Babbage Road, Cambridge CB3 0FS, United Kingdom.

2. CNR-SPIN, c/o University of Salerno, I-84084 Fisciano, Salerno, Italy.

*Corresponding author. Email: angelo.dibernardo@uni-konstanz.de (A.D.B.)
and jjr33@cam.ac.uk (J.W.A.R.)

†Current address: Department of Physics, University of Konstanz, Universitätstraße 10, 78457 Konstanz, Germany.

**This file contains:**

- Supplementary text with:
  - description of the Dirac-like excitation spectrum of quasiparticles for nodal *d*-wave superconductors
  - details about our theoretical model and analysis
  - comparison between nodal exchange coupling and RKKY coupling
  - estimate of $\Delta T_c$ from the theoretical model
- Additional references.
- Supplementary Figures 1-4.

**Dirac-like low-energy excitations in nodal *d*-wave superconductors**

Recent studies have demonstrated the important role played of Dirac-like low-energy excitations in determining the physical properties of a variety of materials including topological insulators[36,37], graphene[38,39], and nodal *d*-wave superconductors[40,41]. In this section, following previous work done on the same topic[42,43], we briefly explain the analogy made in the manuscript between Dirac-like low-energy excitations and the excitation spectrum of quasiparticles in the nodal *d*-wave superconductor (S) YBa$_2$Cu$_3$O$_{7-\delta}$ (YBCO). We also provide details about the formalism used to describe such Dirac-like excitation spectrum for YBCO, which we adapt in the theoretical model reported in the next section of this Supplementary Information.

Conventional superconductors such as Nb and Al are well-described by the Bardeen-Cooper-Schrieffer (BCS) theory of superconductivity[44] and are characterized by an isotropic gapped superconducting order parameter (OP) with *s*-wave symmetry in momentum-space (*k*-space). The fully-gapped OP means that the energy required to excite quasiparticles in *s*-wave superconductors is isotropic in *k*-space.

Unconventional S materials are not fully described by the BCS theory of superconductivity and have an anisotropic OP which is gapless in some points (called 'nodes') of the *k*-space. In these materials, quasiparticles are excited at very low energies, and the quasiparticle excitation spectrum assumes a Dirac-like profile near the nodes, meaning that the energy needed to excite quasiparticles increases linearly with *k* when moving away from a node. Examples of unconventional nodal S materials include the *d*-wave superconductor YBCO, the heavy fermion systems such as CeCoIn$_5$ and URu$_2$Si$_2$ (refs. 45, 46), the nodal non-centrosymmetric superconductors like ThCoC$_2$ (ref. 47), the so-called Weyl superconductors[48], and the A-phase of superfluid $^3$He (ref. 49). While these systems differ in terms of the symmetries and mechanisms that control the formation of their nodal points or lines, they exhibit properties that are directly linked to the physics of Dirac-like low-energy excitations.

To describe the state of a given physical system at low excitation energies, it is certainly necessary to take into account the symmetry of the excitation spectrum. For most metals and semiconductors, for example, low-energy excitations are described by a quadratic Hamiltonian $H = p^2/(2m^*)$, where $p$ and $m^*$ are the electron momentum and effective mass, respectively. For physical systems exhibiting Dirac-like excitations instead, the low-energy description is based on the Dirac Hamiltonian which in two dimensions has the form[43]

$$H = v_{Fx}\,\sigma_x p_x + v_{Fy}\,\sigma_y p_y + m\sigma_z, \quad (S1)$$

where $(\sigma_x, \sigma_y, \sigma_z)$ are the conventional Pauli matrices, $m$ is the mass of the Dirac quasiparticles (the Pauli matrices can describe either the spin or the pseudospin degrees of freedom in a given material), and $(v_{Fx}, v_{Fy})$ the components of the Fermi velocity around the node. In the limit of vanishing Dirac mass (i.e. $m = 0$), the energy spectrum is zero at $(p_x, p_y) = 0$ with a linear dispersion around it, and thus it is fundamentally different from the parabolic dispersion which is encountered in metals or semiconductors. Remarkably, the mixing of positive and negative energy configurations for the two spinor components also occurs in the case of non-vanishing mass, leading to electron- and hole-like states which are correlated and share the same effective mass. The Dirac-like excitation spectrum given by (S1) can be used to describe a variety of systems at low excitation energies[43], although the symmetries and types of spinor modes is system-dependent. These systems include 2D materials such as graphene[50], the surface of topological insulators[51,52] and also nodal *d*-wave S materials[53].

To show how the low-energy excitation spectrum of quasiparticles for a nodal *d*-wave S is equivalent to a Dirac-like excitation spectrum such as that described by the Equation **S1**, we write the Hamiltonian for a *d*-wave superconductor in the basis of the Nambu spinor

$$D^\dagger_{\boldsymbol{k}} = \begin{pmatrix} c^\dagger_{\boldsymbol{k}\uparrow} & c_{-\boldsymbol{k}\downarrow} \end{pmatrix} \quad (S2)$$

as

$$\mathcal{H}_{dw} = \sum_{\boldsymbol{k}} D^\dagger_{\boldsymbol{k}}\, H_{\boldsymbol{k}} D_{\boldsymbol{k}}, \quad (S3)$$

with the matrix $H_{\boldsymbol{k}}$ being given by

$$H_k = (\varepsilon_k - \mu)\tau_z + \Delta_k \tau_x, \tag{S4}$$

where $\tau_i$ (i = x, y, z) are the Pauli matrices in the electron-hole space, $\varepsilon_k$ represents the electronic structure in the S normal state, and $\Delta_k = \Delta_0[\cos k_x - \cos k_y]$ is the $d$-wave gap function. The resulting excitation spectrum for a $d$-wave spin-singlet S can be then obtained by diagonalizing $H_k$ as

$$E_k = \sqrt{(\varepsilon_k - \mu)^2 + |\Delta_k|^2}. \tag{S5}$$

Since the superconducting OP for a $d$-wave S vanishes along the diagonal of the Brillouin zone (BZ) for the values of the momenta $|k_x| = |k_y|$, an electronic dispersion with a closed Fermi surface around either the centre or the corners of the Brillouin zone would lead to points, $\bar{k}$, along the diagonals of the (BZ) which simultaneously satisfy the conditions $\Delta_{\bar{k}} = 0$ and $(\varepsilon_{\bar{k}} - \mu) = 0$ and thus make the excitation spectrum gapless, i.e. $E_{\bar{k}} = 0$. This implies that there exist four nodal points in the BZ which are linked by mirror and inversion symmetries and for which the lowest excitation energy vanishes.

Near the nodal points one can derive the low-energy dispersion by expanding Equation **S1** along the directions perpendicular ($k_\perp$) and parallel ($k_\parallel$) to the Fermi line. Taking the expansion of Equation **S1** around the nodal points up to the lowest order, it can be assumed that the variation of the normal state electronic dispersion is mainly occurring along $k_\perp$, meaning that $(\varepsilon_k - \mu) \sim v_\perp k_\perp$ with $v_\perp = \frac{d\varepsilon_k}{dk_\perp}$. However, since the OP is vanishing along $k_\perp$, the main variation in $\Delta_k$ is due to $k_\parallel$ such as $\Delta_k \sim v_\parallel k_\parallel$ with $v_\perp = \frac{d\Delta_k}{dk_\parallel}$. Here, $v_\perp$ and $v_\parallel$ represent the Fermi and gap velocities, respectively, with ratio $\frac{v_\parallel}{v_\perp}$ typically between 10 and 40 for most cuprates including YBCO with the exception of Tl-based compounds for which $\frac{v_\parallel}{v_\perp}$ is typically larger than 200 (ref. 54). As a result of this expansion, Equation **S4** can be written around the nodal points as

$$H_k \approx v_\perp k_\perp \tau_z + v_\parallel k_\parallel \tau_x, \tag{S6}$$

which has the same form of an anisotropic massless Dirac-like Hamiltonian as that given by Equation **S1**.

**Theoretical model and analysis**
In this section, we provide details on the theoretical analysis carried out to support the experimental results obtained for the PCMO/YBCO/PCMO trilayers. We model the ferromagnetic insulator/superconductor/ferromagnetic insulator (FI$_1$/S/FI$_2$) trilayer with a nodal ($d$-wave) S as a multilayer system with normal to the S/FI interfaces oriented along the $z$-axis (i.e. parallel to the crystallographic $c$-axis), as shown in Figure **S1**.

We start by determining how the spectrum of excitations for a $d$-wave S is modified by the magnetic exchange energy ($E_{ex}$) at each S/FI interface. In particular, we focus on the evolution of the nodal excitations within the 2D Brillouin zone (BZ) as a function of the angle $\theta$ formed by the magnetizations of the two FIs (Fig. **S1**). To simulate a realistic configuration for the trilayer, we compute the u.c.-dependent solution of the superconducting order parameter $\Delta$ that corresponds to the minimum in the free energy of the system ($\mathcal{F}$) and it is obtained by iteratively solving the self-consistent gap Equations. We analyse the spatial profile of the $d$-wave order parameter for any given angle $\theta$ by solving the Bogoliubov-de Gennes Equations for a thin S of volume $V = L_x \times L_y \times L_z$, with $L_x$, $L_y$, $L_z$ being the linear lengths of the trilayer heterostructure along the axes of the crystal lattice. We assume translation invariance in the $xy$-plane and a finite number $n_z$ of unit cells (u.c.) in the S along the $z$-axis (Fig. **S1** shows a schematic illustration of the model for $n_z = 6$ u.c.). The computation is performed using a self-consistent iterative procedure[55] to determine $\Delta$, the $\mathcal{F}$ profile and the resulting ground state with the desired degree of accuracy.

After performing the mean mean-field decoupling between the pairing interaction and $E_{ex}$, we adopt the following Hamiltonian to describe the interaction between superconductivity and magnetism

$$\mathcal{H} = \tfrac{1}{2}\sum_{\boldsymbol{k},(i_z,j_z)} c^\dagger{}_{\boldsymbol{k}}(i_z)\,\hat{H}_{\boldsymbol{k}}(i_z,j_z)\,c_{\boldsymbol{k}}(j_z), \tag{S7}$$

where

$$\hat{H}_{\boldsymbol{k}}(i_z,j_z) = \left(2t_{\parallel}[\cos(k_x)+\cos(k_y)]\delta_{i_z,j_z} + t_c \delta_{i_z,j_{z\pm 1}} + \mu\right)\tau_z\sigma_0 + E_{ex}\tau_z \boldsymbol{M_1}\cdot\boldsymbol{\sigma}\,[\delta_{i_z,1}\delta_{1,j_z}] + E_{ex}\tau_z \boldsymbol{M_2}\cdot\boldsymbol{\sigma}\,[\delta_{i_z,n_z}\delta_{n_z,j_z}] - g\,[F(i_z) f_{\boldsymbol{k}} \tau_y \sigma_y]\delta_{i_z,j_z}, \tag{S8}$$

$$c^\dagger{}_{\boldsymbol{k}}(i_z) = \begin{pmatrix} c^\dagger{}_{\boldsymbol{k}\uparrow}(i_z) & c^\dagger{}_{\boldsymbol{k}\downarrow}(i_z) & c_{-\boldsymbol{k}\uparrow}(i_z) & c_{-\boldsymbol{k}\downarrow}(i_z) \end{pmatrix} \tag{S9}$$

is the spinor associated to the particle-hole electron states at a given u.c. coordinate $i_z$, and $\boldsymbol{\tau},\boldsymbol{\sigma}$ are the Pauli matrices acting in the corresponding subspaces.

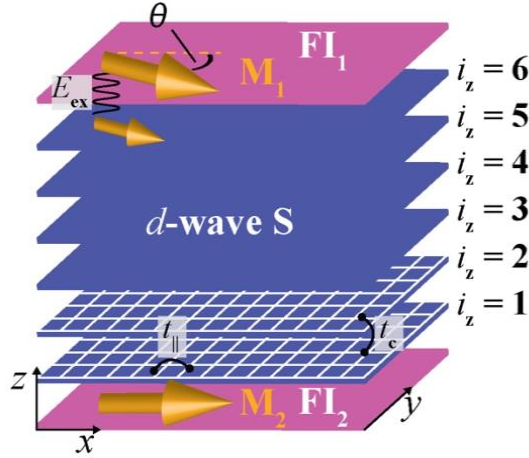

**Fig. S1: Model used for the theoretical analysis.** Schematic illustration of the $FI_1/S/FI_2$ trilayer showing the nearest-neighbour intra- ($t_{\parallel}$) and inter-u.c. ($t_c$) hopping processes, the magnetic exchange energy due to the coupling between the spin polarization in each FI and the S at their interface ($E_{ex}$)), and the misalignment angle $\theta$ formed by the magnetizations of the FI layers.

From Equation **S8** it is clear that the Hamiltonian in Equation **S7** consists of different terms that describe the processes of intra- and inter-u.c. hopping via $t_{\parallel}$ and $t_c$, respectively, the magnetic exchange energy $E_{ex}$ between the magnetizations $\boldsymbol{M_1}$ and $\boldsymbol{M_2}$ and the local electron spin polarization in the S unit cells closer to the interface, as well as the pairing interaction $g$ ($g>0$) in the $d$-wave channel enforced by the $k$-dependent factor $f_{\boldsymbol{k}} = [\cos(k_x) - \cos(k_y)]$. Hereafter, for convenience we assume that $t_{\parallel} = 1$, which sets the overall energy scale. Since the pairs in the S have spin-singlet symmetry, to properly take into account the role play by the misalignment angle θ between $\boldsymbol{M_1}$ and $\boldsymbol{M_2}$, it is sufficient to consider a planar symmetry (i.e. $\boldsymbol{M_1}$ and $\boldsymbol{M_2}$ lay in the $xy$-plane, with $M_1 = m(1,0,0)$ and $M_2 = m(\cos(\theta),\sin(\theta),0)$, $m$ being the amplitude of the local magnetic moment). Since the trilayer is symmetric, we also assume that the strength of $m$ is the same for the two FIs. We verify that asymmetries in the amplitude of $m$ can affect the low-energy electronic structure of the superconductor but they do not qualitatively change the results.

For the unit cells that form the $d$-wave S instead, in our model $F(i_z)$ sets the u.c.-dependent pair correlation amplitude associated to the intra-u.c. spin-singlet $d$-wave symmetry, which is expressed by

$$F(i_z) = \frac{1}{L_x L_y}\sum_{\boldsymbol{k}} f_{\boldsymbol{k}} \langle c^\dagger{}_{\boldsymbol{k}\uparrow}(i_z) c^\dagger{}_{-\boldsymbol{k}\downarrow}(i_z)\rangle, \tag{S10}$$

with $\langle A\rangle$ being the thermal expectation value of a given operator $A$. The gap amplitude Δ is therefore given by $\Delta(i_z) = g\,F(i_z)$. For the averaging in $k$-space, we employ an adaptive interpolating scheme which is suitable to capture the low-energy scales close to the nodal excitations in the Brillouin zone.

We observe that the magnetic coupling is due to the effective exchange $E_{ex}$ at the S/FI interface, arising from charge-transfer processes of $d$-electrons in the transition metals of the Cu-O-Mn bond forming in the manganite/cuprate/manganite trilayer. We also point out that the magnetic modelling can be applied to any interface as it describes the short distance magnetic exchange between magnetically-active atoms. The magnetic interaction is responsible for the formation of a net spin-polarization, which is collinear to the direction of the FI magnetization, within the first S u.c. adjacent to the FI.

Since we are interested in establishing the most energetically-favourable state of the FI/S/FI trilayer as a function of $\theta$, it is useful to introduce the Gibbs free energy $\mathcal{F}$, which that is defined as

$$\mathcal{F} = -\frac{1}{\beta V}\sum_{k,n}\left(\ln\left[e^{-\beta E_{n,k}} + 1\right]\right) + g\frac{1}{L_z}\sum_{i_z}|\Delta(i_z)|^2, \quad (S11)$$

where $\beta = \frac{1}{k_B T}$ and $E_{n,\mathbf{k}}$ indicates the eigenstate of the Bogoliubov-de Gennes Hamiltonian $\mathcal{H}$ at a given $\mathbf{k}$ in the BZ.

Firstly, we describe the modifications in the excitation spectra $E_{n,\mathbf{k}}$ as a function of $\theta$ and $E_{ex}$. A qualitative description can be already obtained assuming a spatially-uniform profile for $\Delta(i_z)$. When $E_{ex}$ is null, the electronic structure of a $d$-wave S consisting of a given number of u.c. ($n_z$) results in $n_z$ point nodes along the diagonals at $k_x = \pm k_y$ of the 2D BZ. Here, each node exhibits a linear dispersion with Dirac-like low-energy gapless excitations. If $E_{ex}$ is non-null instead, a spin-polarization forms in the S due to time-reversal symmetry (TRS) breaking and this results in the absence of Kramers degeneracy. The TRS breaking leads to a splitting of the Dirac cones, meaning that $d$-wave S is converted from a S with $n_z$ nodes at its Fermi surface into a S with $n_z$ Fermi pockets at its Fermi surface. These Fermi pockets have an oval shape since they preferentially extend along a direction that is perpendicular to the diagonals of the 2D BZ, as shown in Fig. **S2**.

The scenario emerging from the evolution of the Fermi surface in the 2D BZ as a function of $\theta$ and $E_{ex}$ allows to gain further insights into the nature of the exchange coupling in these FI/S/FI trilayers. The theoretical observation that the size of the Fermi pockets is tuned by $\theta$ implies existence of a natural competition between the parallel (P) and antiparallel (AP) magnetic configurations of the FIs (Fig. **S2a-c**). The reasons for this competition can be inferred from arguments purely based on symmetry. In the AP state, for example, assuming that the magnetic moments are symmetric at both S/FI interfaces, a restoration of the Kramers degeneracy can be obtained when combining inversion (P) and time-reversal (T) transformations so that PT is again a symmetry for the trilayer system. As a result, by varying $\theta$ the nodal $d$-wave S can be brought back to an electronic configuration with $n_z$ nodes at the Fermi surface when the magnetizations are AP, as explicitly shown in Fig. **S2a-c**. A modification of the nodal spectra is also obtained, for a fixed $\theta$, if the strength of the magnetic exchange $E_{ex}$ varies, with an increase in the Fermi pocket size as $E_{ex}$ increases (Fig. **S2d-f**).

Remarkably, there is a clear trend for the $\theta$-dependence of the ground-state energy associated to any given $k$-point in the BZ, which explains the competition existing between the P ($\theta = 0$) and AP ($\theta = \pi$) configurations at being the overall preferred energy state of the system. To understand such trend, we treat the $\theta$-evolution of the ground-state energy of each $k$-channel as if this were separated from the rest of the BZ, which is a reasonable assumption since the in-plane momentum is a good quantum number. Starting from the Fermi surface configuration in the P case (i.e. $\theta = 0$), the $k$-channels in the 2D BZ can be divided between two regions, namely those laying inside or outside the Fermi pockets.

For a $k$-channel inside a Fermi pocket at $\theta = 0$, the shrinking of the Fermi pocket for an increasing $\theta$ (Fig. **S2a-c**) implies that this mode, which is initially representative of a gapped excitation, becomes progressively gapless (as $\theta$ increases). The $k$-channel hence experiences an energy increase when moving away from $\theta = 0$, meaning that $\theta = 0$ (P state) represents a configuration of energy minimum. As $\theta$ is further increased and the $k$-channel crosses the zero-energy level, however, the parity of the Cooper pairs associated with the u.c. degree of freedom changes. Therefore, an additional shrinking of the Fermi pockets leads to gapping the quasiparticle states close to zero energy, which results in an overall energy decrease leading to a second relative energy minimum for $\theta = \pi$ (AP state).

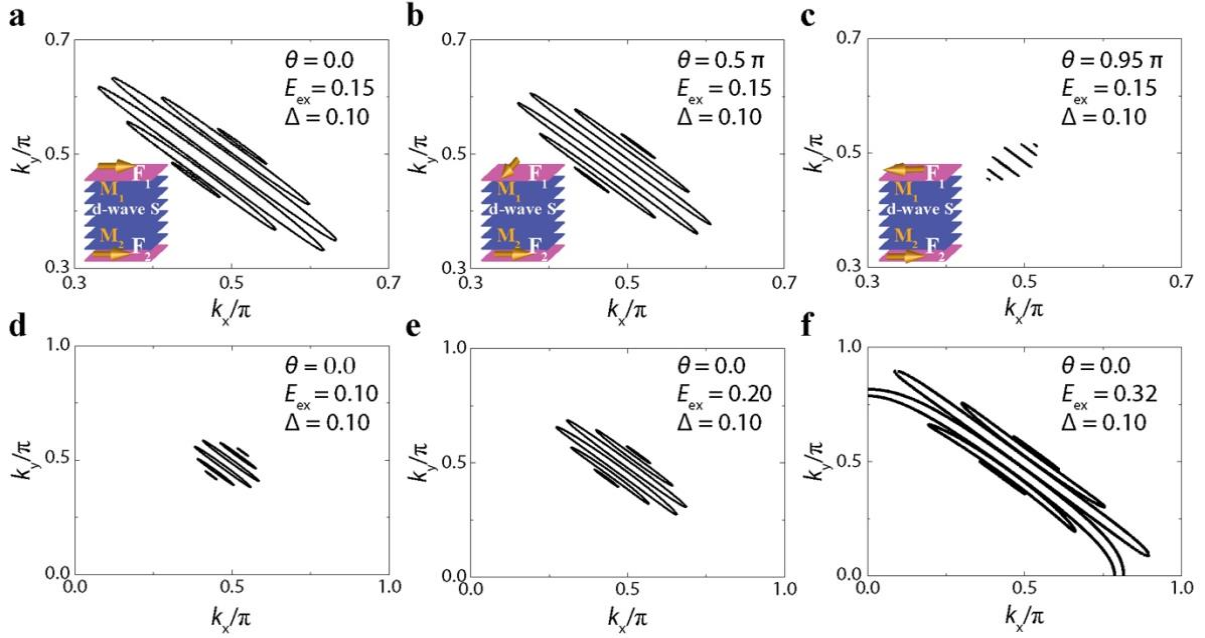

**Fig. S2: Fermi surface for a FI/S/FI trilayer with a nodal ($d$-wave) S. a-f**, Evolution of the nodal excitations in a $d$-wave S consisting of 6 unit cells as a function of the misalignment angle $\theta$ between the magnetizations of the FIs for $\theta = 0$ (**a**), $\theta = \pi/2$ (**b**) and $\theta = 0.95\ \pi$ (**c**), and as a function of the amplitude of the magnetic exchange energy $E_{ex}$ with $\theta = 0$ for $E_{ex} = 0.1$ (**d**), $E_{ex} = 0.2$ (**e**) and $E_{ex} = 0.32$ (**f**).

In summary, depending on the position of the $k$-channel selected, its energy configuration can have minima at $\theta = 0$ (P-state) and/or at $\theta = \pi$ (AP-state). This is also shown in Fig. **3i** (insets) in the main text, where we report the evolution of the free energy for selected $k$-channels in a trilayer with $n_z = 6$ u.c. Since the total energy minimum in the free energy $\mathcal{F}(\theta)$ of the system is calculated by summing up the energy contributions coming from all the $k$-channels in the BZ, the single $k$-channel behaviour explains the intrinsic competition in having either a F-like or AF-like nodal exchange coupling mediated by the $d$-wave S, which we observe experimentally.

We lastly determine the energetically most-favourable spatial profile of the pairing amplitude $\Delta$ by solving the self-consistent Equations for $F(i_z)$. Fig. **S3** shows the results obtained for the representative case with $n_z = 6$ u.c., calculated with different $E_{ex}$ values for both $\theta = 0$ or $\theta = \pi$ and assuming realistic values[56] of the pairing coupling $g = 0.06$ and inter-u.c. hopping $t_c = 0.35$ (in units of $t_{||}$). This choice of parameters corresponds to $\Delta = 10$ meV for $E_{ex} = 0$ and $t_{||} \sim 350$ meV, the latter being a typical d-d hopping amplitude value used for cuprates and oxides in general.

The $\Delta(i_z)$ profiles reported in Fig. **S3a**, **b** exhibit a minimum in the middle of S and an amplitude enhancement at both S/FI interfaces, which occurs for both the P and AP configurations. This behaviour, however, changes for different values of $g$ and $t_c$ than those reported above. For $t_c \sim 0.1$, for example, stable solutions are found for large values of $E_{ex}$ which show the opposite trend from that in Fig. **S3**, meaning a maximum in $\Delta$ in the middle of the S and two minima at the S/FI interfaces. We verify that such differences in the spatial profile of $\Delta$, however, do not affect the peculiar competition between P and AP configurations in the nodal FI/S/FI system.

The $\Delta(i_z)$ curves in Fig. **S3a**, **b** also show a decrease in $\Delta$ as $E_{ex}$ increases, which is obtained for both the P and AP configurations. We also calculate the difference in free energy between P and AP state corresponding to the $\Delta(i_z)$ profiles reported in Fig. **S3a**, **b**. The results shown in Fig. **S3c** indicate that an AP-to-P ground state transition can be obtained as $E_{ex}$ increases, which is consistent with the experimental findings reported in Extended Data Fig. **14** and Extended Data Fig. **15**, where we show that the trilayers always settle in the AP state as the Ca doping in $Pr_{1-x}Ca_xMnO_3$ (and hence $E_{ex}$) is decreased. This observation also confirms that the transition from the AP to the P configuration as $E_{ex}$ is varied is not qualitatively affected if an inhomogeneous $\Delta(i_z)$ is considered in the evaluation of the free energy $\mathcal{F}$. We verify that similar trends for $\Delta(i_z)$ and free energy to those reported in Fig. **S3** are obtained when different values of the pairing interaction $g$, inter-u.c. hopping $t_c$ and number of u.c. $n_z$ are used for the calculations.

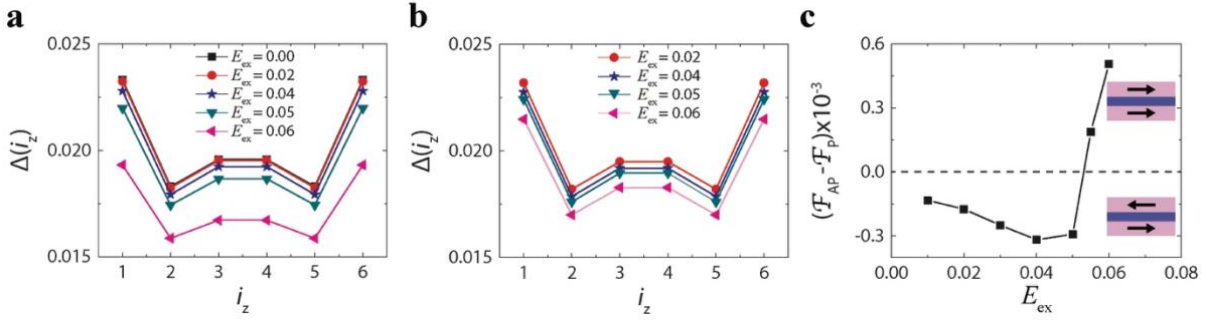

**Fig. S3: Superconducting order parameter and energy ground state of a FI/S/FI trilayer with a nodal (*d*-wave) S. a, b**, Spatial profile of the pairing gap amplitude $\Delta(i_z)$ in a S with $n_z = 6$ u.c. as a function of the magnetic exchange energy $E_{ex}$ at the S/FI interface for parallel with $\theta = 0$ (**a**) and antiparallel with $\theta = \pi$ (**b**) alignment of the magnetizations of the FIs. **c**, Difference in free energy between antiparallel and parallel configurations for the same S in **a** and **b** as a function of $E_{ex}$ showing an AP-to-P transition as $E_{ex}$ is increased (the models in the inset show the alignment of the FIs in the energy ground state). The energy values in all the panels are normalized to $t_{\parallel}$.

Fig. **S4a** shows that the global $\mathcal{F}(\theta)$ of the FI/S/FI trilayer ($n_z = 8$ u.c.) can exhibit two relative minima at $\theta = 0$ (P state) to $\theta = \pi$ (AP state) for some choices of the model parameters. In such situation, $\mathcal{F}(\theta)$ reaches a maximum value between the two local minima as $\theta$ is varied between 0 and $\pi$. This result is consistent with the experimental observation of a maximum in $R(H)$ measured close to $H = 0$ (i.e. for a magnetically-inhomogeneous state of the system) and associated with two local minima in $R(H)$ in the AP and P state (see Fig. **2j, k** and Extended Data Fig. **8b** of the manuscript). The $\mathcal{F}(\theta)$ profiles in Fig. **S4a** calculated for different $E_{ex}$ values also suggest that the $\mathcal{F}(\theta)$ maximum shifts in $\theta$ depending on $E_{ex}$.

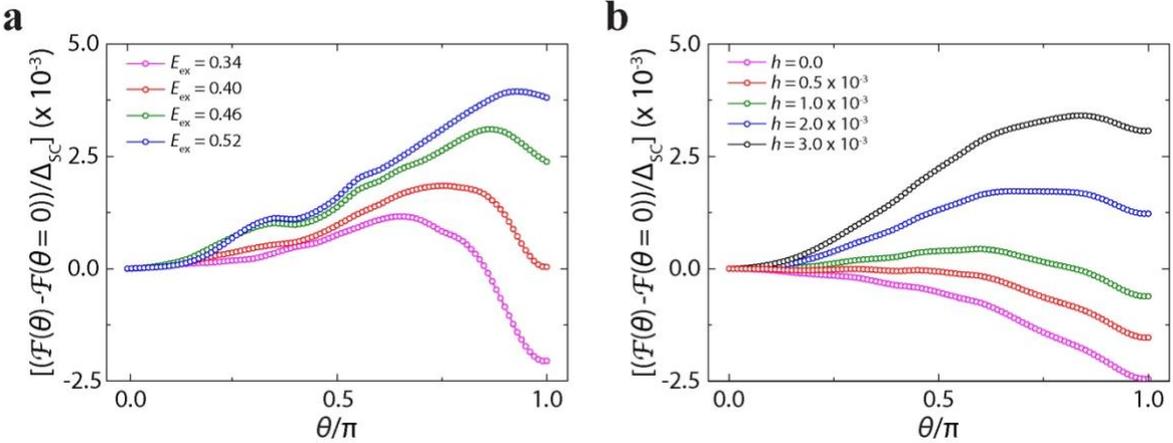

**Fig. S4: Evolution of the energy ground state of a FI/S/FI trilayer with a nodal (*d*-wave) S as a function of exchange energy and applied magnetic field. a,b**, Variation of the global free energy $\mathcal{F}(\theta)$ of the FI/S/FI system ($n_z = 8$ u.c.) with the misalignment angle $\theta$ between the FIs, for increasing values of the exchange energy $E_{ex}$ (**a**) in zero applied magnetic field and for increasing values of the applied field (modelled through the parameter $h$) at fixed $E_{ex}$ (**b**). $\mathcal{F}(\theta)$ is plotted after subtracting its value at $\theta = 0°$ (P state) and normalising the difference by the superconducting gap amplitude $\Delta$. For the calculation of the profiles in (**a**) we assume $t_c = 0.45$, $h = 0$ and for those in (**b**) $t_c = 0.25$ and $E_{ex} = 0.35$, with all the parameters being expressed in units of $t_{\parallel}$.

Finally, we simulate the effect of an applied magnetic field on $\mathcal{F}(\theta)$ by introducing a Zeeman term in the Hamiltonian in Equation **S8** through a parameter $h$ (in units of $t_{\parallel}$) which couples to the electronic states in S as $E_{ex}$. The results of these simulations are reported in Fig. **S4b** for $n_z = 8$ u.c., which shows that, as $h$ increases, the absolute minimum in $\mathcal{F}(\theta)$ changes from $\theta = \pi$ (AP state) to $\theta = 0$ (P state), whilst simultaneously a maximum in $\mathcal{F}(\theta)$ arises at $\theta$ corresponding to a non-collinear configuration of the FIs (i.e. for $0 < \theta < \pi$). We also verify that different values of $t_c$ and $E_{ex}$ only result in changes in the length scale at which the described evolution of $\mathcal{F}(\theta)$ with $h$ occurs, but they do not alter qualitatively the trend reported in Fig. **S4b**.

**Comparison between nodal exchange coupling and RKKY coupling**

The oscillations in sign and amplitude of $\Delta T_c = T_c(\text{P}) - T_c(\text{AP})$ with superconductor thickness $d_S$, which we observe in our trilayers, may suggest that the nodal exchange coupling responsible for the oscillations shares some similarities with a Ruderman-Kittel-Kasuya-Yosida (RKKY) type of coupling generated between magnetic moments (localized spins) embedded in a metallic host[34,57,58]. In this section, we discuss aspects that differentiate nodal exchange coupling in FI/S/FI trilayers to RKKY coupling.

The RKKY interaction describes a mechanism of spin exchange where a localized magnetic moment induces a spin polarization in the surrounding conduction electrons, which mediates the coupling with a second magnetic moment at a given distance ($r$) from the first moment. The key signature of this RKKY mechanism is the oscillatory sign of the exchange interaction as a function of $r$.

When considering the effect of superconductivity on RKKY coupling within the lowest-order perturbation theory, the effective exchange interaction becomes antiferromagnetic-like if $r \gtrsim \xi_s$ ($\xi_s$ being the superconductor coherence length), while for distances $r < \xi_s$ the standard RKKY represents the main contribution to the magnetic coupling together with a weak antiferromagnetic correction due to the superconducting pairing[59,60,61]. Such perturbation approach, however, neglects the formation of magnetic bound states, which are localized electronic states around the magnetic moments[62,63,64]. Depending on the energy of the bound states within the superconducting gap and on the degree of electronic hybridization, the effective spin exchange can be completely modified so that an antiferromagnetic exchange becomes dominant even for distances $r < \xi_{sc}$ (ref. 65).

In our FI/S/FI trilayers, while the magnetizations of the FIs can couple, as a for a typical RKKY-like scenario, by interacting with the spin-polarised electrons existing close the S/FI interfaces, there are additional mechanisms that contribute to the overall exchange coupling. Firstly, due to S being nodal ($d$-wave), at any given channel in momentum space ($k$-channel), the effective coherence length can satisfy either $r < \xi_s$ and $r \gtrsim \xi_s$, with $r$ corresponding to the S thickness $d_S$ (i.e. to the distance between the moments at FI/S interfaces that couple).

In addition, in our system the local magnetic exchange $E_{ex}$ at the FI/S interfaces is larger than the superconducting gap amplitude $\Delta$ and comparable in magnitude with the interlayer hopping $t_c$. This observation implies that modifications to the electronic states cannot be neglected, and therefore a conventional perturbation approach would not be sufficient to capture the qualitative and quantitative features of our effective exchange coupling mechanism.

Another important difference compared to the conventional RKKY scenario is that the electronic states along the direction perpendicular to the layers (i.e. $z$ direction in our model) are confined by the presence of the FIs. In addition, due to the anisotropic nature of the superconducting order parameter in the nodal S, $t_c$ is smaller compared to intra-layer hopping $t_{||}$. It also evident that reflections at the FI/S interfaces must be considered because of the itinerant character of the electrons along the $z$ direction. Furthermore, typical momenta along $z$ are of the order of $d_S^{-1}$. All these factors imply that the different $k$-channels introduce functions with odd- and even-spatial symmetries which combine to generate the overall exchange coupling.

The change in sign of the nodal exchange coupling is controlled by the amplitude of $E_{ex}$ at FI/S interfaces, which can induce a transition from a purely-antiferromagnetic exchange to a sign-changing oscillatory exchange. The confinement and the $k$-dependence of $\xi_s$ for a $d$-wave S allow then to have competing channels which contribute to the overall exchange coupling as either short- or long-ranged exchange coupling terms with non-equal amplitudes and signs. The resulting scenario is that both ferromagnetic and antiferromagnetic effective exchanges can dominate depending on $d_S$ and on the strength of $E_{ex}$ at the FI/S interfaces.

In summary, both conventional RKKY-like coupling and exchange coupling due to hybridization of spin-polarized electronic bound states contribute to the effective (nodal) exchange coupling between FIs in our FI/S/FI system. Taking this into account, we can state that the oscillatory behavior of the nodal exchange coupling in our trilayers has a characteristic length scale that is of the order of a few unit cells as shown in Fig. **3j** of the manuscript.

**$\Delta T_c$ variation in a FI/S/FI system with a nodal *d*-wave S**
An estimate of the $\Delta T_c$ values predicted by our theoretical model can be obtained by observing that the variation of the $T_c$ between P and AP states has to follow that one of the free energy $\mathcal{F}$ of the spin-valve between the same states (i.e. $\Delta \mathcal{F} = |\mathcal{F}_P - \mathcal{F}_{AP}|$).

Since $\mathcal{F}$ exhibits two minima corresponding to P and AP states and since the energy splitting between these two minima is related to $E_{ex}$ in the presence of non-zero superconducting gap, the energy difference between the P and AP configurations must also set the difference in $T_c$ between the two states (i.e. $\Delta T_c$). This can be understood, for example, assuming that the absolute minimum in $\mathcal{F}$ at low temperatures corresponds to the AP state and that $\mathcal{F}_{AP}$ is separated by $\Delta F$ from $\mathcal{F}_P$. In this case, as the FI/S/FI system approaches $T_c$ and undergoes a transition from the superconducting into the normal-state, the presence of a non-null $\Delta \mathcal{F}$ makes the normal-state already more-energetically favourable than the P state, meaning that $T_c$ (AP) > $T_c$ (P).

It is also reasonable to expect that $\Delta T_c$ is proportional to $\Delta \mathcal{F}$, since the thermal distribution of the superconducting excitations does not change significantly between P and AP states. Deviations from this condition can occur only when the P and AP states are very close in energy or almost degenerate. Such correlation existing between $\Delta T_c$ and $\Delta \mathcal{F}$ is also confirmed by the similar oscillatory behaviour of both quantities as a function of the S thickness $d_S$.

Although the nodal exchange includes several effects related to the modification of the electronic states close to the interface and the presence of nodal excitations in the *d*-wave S, a qualitative estimate of the magnitude of $\Delta T_c$ can be obtained by observing that $\Delta \mathcal{F}$ is proportional to the effective magnetic exchange $J_{ex}$ between the F magnetizations. Hence, considering the various electronic processes which are involved, one can argue that the expression holding for the superconducting mediated-exchange between magnetic moments[65] $J_{ex}$ can be used to estimate for $\Delta T_c$, once this expression is adapted to the case of a layered thin superconducting film. Therefore, $J_{ex}$ and $\Delta T_c$ can be estimated from

$$\Delta T_c \sim J_{ex} \sim \frac{t_c \left(\frac{E_{ex}}{t_\parallel}\right)^2}{(k_{Fz} d_S)^3} + \frac{\Delta_{SC} \left(\frac{E_{ex}}{t_\parallel}\right)^2}{(k_{Fz} d_S)^2}. \quad (S12)$$

with $k_{Fz}$ being the effective Fermi wave vector for the electron propagation along the out-of-plane direction. Assuming values of $k_{Fz} d_S$ in the range [2-4] and of $\frac{E_{ex}}{t_\parallel}$ in the range $[10^{-2}; 10^{-1}]$, and using the typical amplitudes for the superconducting gap of YBCO $\Delta_{SC}$ (~ 10 meV) and the inter-layer hopping $t_\parallel$ (~ 350 meV), one obtains that $\Delta T_c$ is of the order of a few kelvins.

We also point out that the amplitude of $\Delta \mathcal{F}$ obtained from our numerical computation is also associated to a magnetic exchange energy, and therefore to a $\Delta T_c$ value which is of the order of a few kelvins, consistently with the experimental results reported in Fig. **3g** of the main text.

## Supplementary references

36. Qu, D. X., Hor, Y. S., Xiong, J., Cava, R. J. & Ong, N. P. Quantum oscillations and Hall anomaly of surface states in the topological insulator Bi$_2$Te$_3$. *Science* **329**, 821-824 (2010).
37. Checkelsky, J. G., Ye, J., Onose, Y., Iwasa, Y. & Tokura, Y. Dirac-fermion-mediated ferromagnetism in a topological insulator. *Nature Phys.* **8**, 729-733 (2012).
38. Novoselov, K. S., Geim, A. K., Morozov, S. V., Jiang, D., Zhang, Y., Dubonos, S. V., Grigorieva, I. V. & Firsov, A. A. Electric field effect in atomically thin carbon films. *Science* **306**, 666–669 (2004).
39. Kane, C. L. & Mele, E. J. Quantum spin Hall effect in graphene. *Phys. Rev. Lett.* **95**, 226801 (2005).
40. Proust, C., Boaknin, E., Hill, R. W., Taillefer, L. & Mackenzie, A. P. Heat transport in a strongly overdoped cuprate: Fermi liquid in a pure *d*-wave BCS superconductor. *Phys. Rev. Lett.* **89**, 147003 (2002).